\documentclass[]{article}
\usepackage{multicol}
\usepackage{graphicx}
\usepackage{amssymb}
\usepackage{epsfig}
\usepackage[greek,english]{babel}
\usepackage[iso-8859-7]{inputenc}
\usepackage{subcaption}
\usepackage{amsmath}
\usepackage{float}
\usepackage{verbatim}
\usepackage{wrapfig}
\usepackage{textcomp}
\usepackage{color}
\usepackage{sidecap}
\usepackage{cancel}
\usepackage{amssymb}
\usepackage{kerkis}
\usepackage{youngtab}
\usepackage{stackrel}
\usepackage[toc,page]{appendix}
\usepackage{multirow}
\usepackage{bm}
\usepackage{cite}
\usepackage[absolute]{textpos}
\usepackage[stable]{footmisc}
\newcommand{\by}{by }
\newcommand{\thee}{the }
\newcommand{\as }{as }
\newcommand{\andd }{and }
\newcommand{\newc}{\newcommand}
\newc{\ra}{\rightarrow}
\newc{\lra}{\leftrightarrow}
\newc{\ov}{\overline}
\newc{\pa}{\partial}

\newc{\be}{\begin{equation}}
\newc{\ba}{\begin{eqnarray}}
\newc{\ea}{\end{eqnarray}}
\newc{\n}{\nu}
\newc{\D}{\Delta}

\newc{\la}{\lambda}
\newc{\e}{\epsilon}
\newc{\nn}{\nonumber}

\def\bea{\begin{eqnarray}}
\def\eea{\end{eqnarray}}
\begin{document}
\title{Higher-Dimensional Unified Theories with continuous and fuzzy coset
spaces as extra dimensions\thanks{Based on a talk given \by G.Z. as invited main speaker at the $11-$th International Workshop on "Lie Theory and Its Applications in Physics" (LT-11), 15-21, June 2015, Varna, Bulgaria", to be published in Springer.}}

\author{G. Manolakos\textsuperscript{1},\,G. Zoupanos\textsuperscript{1,2}}\date{}
\maketitle
\begin{center}
\emph{E-mails: gmanol@central.ntua.gr\,, George.Zoupanos@cern.ch }
\end{center}

\begin{center}
\itshape\textsuperscript{1}Physics Department, National Technical
University,\\ Zografou Campus, GR-15780 Athens, Greece\\
\itshape\textsuperscript{2}Institut f{\"u}r Theoretische Physik
Universit{\"a}t \\ Heidelberg Philosophenweg 16, D-69120 Heidelberg
\end{center}
\vspace{0.1cm} \emph{Keywords}: higher-dimensional theories, coset
space dimensional reduction, fuzzy spheres, orbifold projection

\begin{abstract}
  We first briefly review \thee Coset Space Dimensional Reduction (CSDR)
programme and present \thee results of the best model so far, based on \thee
$\mathcal{N} = 1$, $d = 10$, $E_8$ gauge theory reduced over
\thee nearly-K{\"a}hler manifold $SU(3)/U(1)\times U(1)$. Then, we present the adjustment of the CSDR programme in the case that the extra dimensions are considered to be fuzzy coset spaces and then, the best model constructed in this framework, too, which is \thee trinification GUT, $SU(3)^3$.
\end{abstract}
\section{Introduction}

During \thee last decades, unification of \thee fundamental interactions has focused \thee interest of theoretical physicists. This has led to \thee rise of very interesting \andd well-established approaches. Important \andd appealing are \thee ones that elaborate extra dimensions. A consistent framework in this approach is superstring theories \cite{green-scwarz-witten} with \thee Heterotic String \cite{gross-harvey} (defined in ten dimensions) being the most promising, due to the possibility that in principle could lead to experimentally testable predictions. More specifically, the compactification of \thee $10-$dimensional spacetime \andd \thee dimensional reduction of \thee $E_8\times E_8$ initial gauge theory lead to phenomenologically interesting Grand Unified theories (GUTs), containing \thee SM gauge group.

A few years before \thee development of \thee superstring theories, another important framework aiming at \thee same direction was employed, that is \thee dimensional reduction of higher-dimensional gauge theories. Pioneers in this field were Forgacs-Manton and Scherk-Schwartz studying \thee Coset Space Dimensional Reduction (CSDR) \cite{forgacs-manton,kapetanakis-zoupanos, kubyshin-mourao} \andd Scherk-Schwarz group manifold reduction \cite{scherk-schwarz}, respectively. In both of these approaches, \thee higher-dimensional gauge fields are unifying \thee gauge \andd scalar fields, while the $4-$dimensional theory contains the surviving components after the procedure of the dimensional reduction. Moreover, in \thee CSDR scheme, the inclusion of fermionic fields in the initial theory leads to Yukawa couplings in the $4-$dimensional theory. Furthermore, upgrading the higher-dimensional gauge theory to $\mathcal{N}=1$ supersymmetric, i.e. grouping the gauge and fermionic fields of the theory into the same vector supermultiplet, is a way to unify further the fields of the initial theory, in certain dimensions. A very remarkable achievement of the CSDR scheme is the possibility of obtaining chiral theories in four dimensions \cite{mantonns,Chapline-Slansky}.

The above context of \thee CSDR adopted some very welcome suggestions coming from \thee superstring theories (specifically from \thee Heterotic String \cite{gross-harvey}), that is \thee dimensions of the space-time and \thee gauge group of \thee higher-dimensional supersymmetric theory. In addition, taking into account \thee fact that \thee superstring theories are consistent only in ten dimensions, \thee following important issues have to be addressed, (a) distinguish \thee extra dimensions from \thee four observable ones \by considering an appropriate compactification of \thee metric and (b) determine \thee resulting $4-$dimensional theory. Additionally, a suitable choice of \thee compactification manifolds could result into $\mathcal{N}=1$ supersymmetry, aiming for a chance to be led to realistic GUTs.

Aiming at \thee preservation of an $\mathcal{N}=1$ supersymmetry after \thee dimensional reduction, Calabi-Yau (CY) spaces serve as suitable compact internal manifolds \cite{Candelas}. However, \thee emergence of \thee moduli stabilization problem, led to \thee study of flux compactification, in \thee context of which a wider class of internal spaces, called manifolds with $SU(3)-$structure, was suggested. In this class of manifolds, a non-vanishing, globally defined spinor is admitted. This spinor is covariantly constant with respect to a connection with torsion, versus \thee CY case, where \thee spinor is constant with respect to \thee Levi-Civita connection. Here, we consider \thee nearly-K{\"a}hler manifolds, that is an interesting class of $SU(3)-$structure manifolds \cite{cardoso-curio, Irges-Zoupanos, i-z, Butruille}. The class of homogeneous nearly-K{\"a}hler manifolds in six dimensions consists of the non-symmetric coset spaces $G_2/SU(3)$,
$Sp(4)/(SU(2)\times U(1))_{non-max}$, $SU(3)/U(1)\times U(1)$ and
the group manifold $SU(2)\times SU(2)$ \cite{Butruille} (see also
\cite{Irges-Zoupanos, i-z, cardoso-curio}). It is worth mentioning that $4-$dimensional theories which are obtained after \thee
dimensional reduction of a $10-$dimensional $\mathcal{N}=1$
supersymmetric gauge theory over non-symmetric coset spaces, contain
supersymmetry breaking terms \cite{manousselis-zoupanos},
\cite{manousselis-zoupanos2}, contrary to CY spaces.

Another very interesting framework which admits a description of physics at \thee Planck scale is non-commutative geometry \cite{connes} - \cite{fuzzy}. Regularizing quantum field theories, or even better, building finite ones are \thee features that render it as a promising framework. On \thee other hand, the construction of quantum field theories on non-commutative spaces is a difficult task and, furthermore, problematic ultraviolet features have emerged \cite{filk} (see also \cite{grosse-wulkenhaar}
and \cite{grosse-steinacker}. However, non-commutative geometry is an appropriate framework to accommodate particle models with non-commutative gauge theories \cite{connes-lott} (see also \cite{martin-bondia,
dubois-madore-kerner, madorejz}).

It is remarkable that \thee two frameworks (superstring theories \andd non-commutative geometry) found contact, after \thee realization that, in M-theory and open String theory, \thee effective physics on D-branes can be described \by a non-commutative gauge theory \cite{connes-douglas-schwarz, seiberg-witten}, if a non-vanishing background antisymmetric field is present. Moreover, \thee type IIB superstring theory (\andd others related with type IIB with certain dualities) in its conjectured non-perturbative formulation as a matrix model \cite{ishibasi-kawai}, is a non-commutative theory. In \thee framework of non-commutative geometry, Seiberg and Witten \cite{seiberg-witten} contributed the most with their study (map between commutative \andd non-commutative gauge theories) based on which notable developments \cite{chaichian, jurco} were achieved \andd afterwards a non-commutative version of \thee SM was constructed \cite{camlet}. Unfortunately, such extensions fail to solve \thee main problem of \thee SM, which is \thee presence of many free parameters.

A very interesting development in \thee framework of \thee non-commutative geometry is \thee programme in which \thee extra dimensions of higher-dimensional theories are considered to be non-commutative (fuzzy) \cite{aschieri-madore-manousselis-zoupanos,
aschieri-grammatikopoulos, steinacker-zoupanos,
chatzistavrakidis-steinacker-zoupanos,fuzzy}. This programme overcomes \thee ultraviolet/infrared problematic behaviours of theories defined in non-commutative spaces. A very welcome feature of such theories is that they are renormalizable, versus all known higher-dimensional theories. This aspect of \thee theory was examined from \thee $4-$dimensional point of view too, using spontaneous symmetry breakings which mimic \thee results of \thee dimensional reduction of a higher-dimensional gauge theory with non-commutative (fuzzy) extra dimensions. In addition, another interesting feature is that in theories constructed in this programme, there is an option of choosing \thee initial higher-dimensional gauge theory to be abelian. Then, non-abelian gauge theories result in lower dimensions in \thee process of \thee dimensional reduction over fuzzy coset spaces. Finally, \thee important problem of chirality in this framework has been addressed \by applying an orbifold projection on a $\mathcal{N}=4$ SYM theory. After \thee orbifolding, \thee resulting theory is an $\mathcal{N}=1$ supersymmetric, chiral $SU(3)^3$.

\section{The coset space dimensional reduction of a $D-$dimensional YMD Lagrangian}

An obvious \andd crude way to realize a dimensional reduction of a higher-dimensional gauge theory is to demand that all \thee fields of \thee theory are independent of \thee extra coordinates (trivial reduction) \andd therefore \thee Lagrangian is independent, too. A much more elegant way is to allow for a non-trivial dependence considering that a symmetry transformation on \thee fields \by an element that belongs in the isometry group $S$ of \thee compact coset space $B=S/R$ formed \by \thee extra dimensions is a gauge transformation (symmetric fields). Therefore, \thee a priori consideration of \thee Lagrangian as gauge invariant, renders it independent of \thee extra coordinates. The above way of getting rid of \thee extra dimensions is \thee basic concept of \thee CSDR scheme \cite{forgacs-manton,
kapetanakis-zoupanos, kubyshin-mourao}.

Let us now consider the action of the $D-$dimensional YM theory with gauge symmetry $G$, coupled to fermions defined on $M^D$ with metric $g^{MN}$
\begin{align}
A&=\int
d^4xd^dy\sqrt{-g}\left[-\frac{1}{4}Tr(F_{MN}F_{K\Lambda})g^{MK}g^{N\Lambda}+\frac{i}{2}\bar
\psi \Gamma^MD_{M}\psi\right]\,,
\end{align}
where $D_M=\pa_M-\theta_M-A_M$, with $\theta_M=\dfrac{1}{2}\theta_{MN\Lambda}\Sigma^{N\Lambda}$ \thee spin connection of $M^D$ and $F_{MN}=\pa_M A_N-\pa_NA_M-[A_M,A_N]$, where $M, \mathcal{N}=1\ldots D$ and $A_M$, $\psi$ are
$D$-dimensional symmetric fields. The fermions can be accommodated in any representation $F$ of $G$, unless an additional symmetry, e.g. supersymmetry, is considered.

Let $\xi^{\alpha}_{A}, (A=1,...,dimS$ and $\alpha =dimR+1,...,dimS$ the curved index$)$ be \thee Killing vectors which generate \thee symmetries of $S/R$ \andd $W_A$,
\thee gauge transformation associated with $\xi_A$. The
following constraint equations for scalar $\phi$, vector
$A_{\alpha}$ \andd spinor $\psi$ fields on $S/R$, derive from \thee
definition of \thee symmetric fields, that is \thee $S-$transformations of \thee fields are gauge transformations
\begin{gather}
\delta_A\phi=\xi^{\alpha}_{A}\partial_\alpha\phi=D(W_A)\phi,\label{ena} \\
\delta_AA_{\alpha}=\xi^{\beta}_{A}\partial_{\beta}A_{\alpha}+\partial_{\alpha}\xi^{\beta}_{A}A_{\beta}=\partial_{\alpha}W_A-[W_A,A_{\alpha}], \label{duo}\\
\delta_A\psi=\xi^{\alpha}_{A}\partial_\alpha\psi-\frac{1}{2}G_{Abc}\Sigma^{bc}\psi=D(W_A)\psi\,,
\label{tria}
\end{gather}
where $W_A$ depend only on internal coordinates $y$ and $D(W_A)$
represents a gauge transformation in \thee corresponding representation
where \thee fields belong. Solving \thee above constraints \eqref{ena}-\eqref{tria}, we result with \cite{forgacs-manton, kapetanakis-zoupanos} \thee unconstrained $4-$dimensional fields, as well as with \thee remaining $4-$dimensional gauge symmetry.

We proceed \by analysing \thee constraints on \thee fields in \thee theory. We start with \thee gauge field $A_M$ on $M_D$, which splits into its components as $(A_\mu, A_\alpha)$ corresponding to $M^4$ \andd $S/R$, respectively. Solving \thee
corresponding constraint, \eqref{duo}, we obtain \thee following information: First, \thee $4-$dimensional gauge field, $A_\mu$ is completely independent of \thee coset space coordinates and second, \thee $4-$dimensional gauge fields commute with \thee generators of \thee subgroup $R$ in $G$. This means that \thee surviving gauge symmetry, $H$, is \thee subgroup of $G$ that commutes with $R$, that is \thee centralizer of $R$ in $G$, i.e. $H=C_G(R_G)$. The $A_\alpha(x,y)\equiv\phi_\alpha(x,y)$, transform as scalars in \thee $4-$dimensional theory \andd $\phi_\alpha(x,y)$ act as interwining operators connecting induced representations of $R$ acting on $G$ and $S/R$. In order to find \thee representation in which \thee scalars are accommodated in \thee $4-$dimensional theory, we have to decompose $G$ according to \thee embedding
\begin{equation}
G\supset R_G\times H\,,\qquad
adjG=(adjR,1)+(1,adjH)+\sum(r_i,h_i)\,,
\end{equation}
and $S$ under $R$
\begin{align}
S\supset R\,,\qquad
adjS=adjR+\sum s_i\,.
\end{align}
Therefore, we conclude that for every pair $r_i, s_i$, where $r_i$ \andd $s_i$ are identical irreducible representations of $R$, there remains a scalar (Higgs) multiplet which transforms under \thee representation $h_i$ of $H$. All other scalar fields vanish.

As far as \thee spinors are concerned \cite{kapetanakis-zoupanos,
Wetterich-Palla, mantonns,Chapline-Slansky}, \thee analysis of \thee corresponding constraint, \eqref{tria}, is quite similar. Again, solving \thee  constraint, one finds that \thee spinors in \thee $4-$dimensional theory are independent of \thee coset coordinates and act as interwining operators connecting induced  representations of $R$ in $SO(d)$ and in $G$. In order to obtain \thee representation of $H$, where \thee fermions are accommodated in \thee resulting $4-$dimensional theory, one has to decompose \thee initial representation $F$ of $G$ under \thee $R_G\times H$,
\begin{equation}
 G\supset R_G\times H\,,\qquad F=\sum (r_i,h_i),
\end{equation}
\andd \thee spinor of $SO(d)$ under $R$
\begin{equation}
SO(d)\supset R\,,\qquad \sigma_d=\sum \sigma _j\,.
\end{equation}
Therefore, for each pair $r_i$ and $\sigma_i$, where $r_i$ and $\sigma_i$ are identical irreducible representations, there exists a multiplet, $h_i$ of spinor fields in \thee $4-$dimensional theory. As for \thee chirality of \thee surviving fermions, if one begins with Dirac fermions in \thee higher-dimensional theory it is impossible to result with chiral fermions in \thee $4-$dimensional theory. Further requirements have to be imposed in order to result with chiral fermions in  \thee  $4-$dimensional theory. Indeed, imposing  \thee  Weyl condition in  \thee  chiral representations of an even higher-dimensional initial theory, one is led to a chiral theory in four dimensions. This is not  \thee  case in an odd higher-dimensional initial theory, in which Weyl condition cannot be imposed. The most interesting case is  \thee  $D=2n+2$ even higher dimensional initial theory, in which starting with fermions in  \thee  adjoint representation  \thee  Weyl condition leads to two sets of chiral fermions with  \thee  same quantum numbers under $H$ of  \thee  $4-$dimensional theory. This doubling of  \thee  fermionic spectrum can be eliminated after imposing  \thee  Majorana condition. The two conditions are compatible when $D=4n+2$, which is  \thee  case of our interest.

 Now, let us move on and determine  \thee  $4-$dimensional effective action. The first and very important step is to compactify  \thee  space $M^D$ to $M^4\times S/R$, with $S/R$ a compact coset space. After  \thee  compactification,  \thee  metric will be transformed to
\begin{equation}
g^{MN}=\left(
         \begin{array}{cc}
           \eta^{\mu \nu} & 0 \\
           0 & -g^{ab} \\
         \end{array}
       \right)\,,\label{metric_comp}
\end{equation}
where $\eta^{\mu \nu}=\text{diag}(1,-1,-1,-1)$ and $g^{ab}$ is  \thee  metric of  \thee  coset. Inserting  \thee  \eqref{metric_comp} into  \thee  initial action and taking into account \thee  constraints of  \thee  fields, we obtain
\begin{align}
A&=C\int d^4x\left[-\frac{1}{4}F_{\mu \nu}^tF^{t\mu
\nu}+\frac{1}{2}(D_\mu \phi _\alpha)^t(D^\mu \phi
^\alpha)^t+V(\phi)+\frac{i}{2}\bar \psi
\Gamma^{\mu}D_{\mu}\psi-\frac{i}{2}\bar \psi
\Gamma^{a}D_{a}\psi\right]\,,\label{tessera}
\end{align}
where $D_\mu = \partial_{\mu}-A_\mu$ and $D_a =
\partial_{a}-\theta_a-\phi_a$, with $\theta_a =
\frac{1}{2}\theta_{abc}\Sigma^{bc}$  \thee  connection of  \thee  space and
$C$ is  \thee  volume of  \thee  space. The potential $V(\phi)$ is
given \by  \thee  following expression
\begin{align}
V(\phi)=-\frac{1}{4}g^{ac}g^{bd}Tr(f^C_{ab} \phi_C -[\phi_a,
\phi_b])(f^D_{cd} \phi_D - [\phi_c, \phi_d]),
\end{align}
where, $A=1,...,dimS$ and $f$'s are  \thee  structure constants
appearing in  \thee  commutators of  \thee  Lie algebra of $S$. Considering  \thee  constraints of  \thee  fields, \eqref{ena}-\eqref{duo}, one finds that  \thee  scalar fields $\phi_a$ have to obey  \thee  following equation:
\begin{align}
f^D_{ai}\phi_D-[\phi_a, \phi_i]=0\,, \label{tessera_ena}
\end{align}
where  \thee  $\phi_i$ are  \thee  generators of $R_G$. Consequently, some fields will be filtered out, while others will survive  \thee  reduction and will be identified as  \thee  genuine Higgs fields. The potential $V(\phi)$, written down in terms of  \thee  surviving scalars (\thee  Higgs fields), is a quartic polynomial which is invariant under  \thee  $4-$dimensional gauge group, $H$. Then, it follows  \thee  determination of  \thee  minimum of  \thee  potential and  \thee  finding of  \thee  remaining gauge symmetry of  \thee  vacuum \cite{Harnad,Chapline-Manton,Farakos-Koutsoumbas}. In general, this is a rather difficult procedure. However, there is a case in which one could obtain  \thee  result of  \thee  spontaneous breaking of  \thee  gauge group $H$ very easily, whether  \thee  following criterion is satisfied. Whenever $S$ has an isomorphic image $S_G$ in $G$, then  \thee  $4-$dimenisonal gauge group $H$ breaks spontaneously to a subgroup $K$, where $K$ is  \thee  centralizer of $S_G$ in  \thee  gauge group of  \thee  initial, higher-dimensional, theory, $G$ \cite{kapetanakis-zoupanos,Harnad,Chapline-Manton,Farakos-Koutsoumbas}. This can be illustrated in  \thee  following scheme,
\begin{align}
  G\supset &\,S_G\times K\nonumber\\
  &\,\cup\quad\,\,\,\cap    \nonumber\\
  G\supset&\,R_G\times H
\end{align}
In addition,  \thee  potential of  \thee  resulting $4-$dimensional gauge theory is always of spontaneous symmetry breaking form, when  \thee  coset space is symmetric\footnote{A coset space is called symmetric when $f_{ab}^c=0$}. A negative result in this case is that, after  \thee  dimensional reduction,  \thee  fermions end up being supermassive, as in  \thee  Kaluza-Klein theory.

Let us now summarize a few results coming from  \thee
dimensional reduction of  \thee  $\mathcal{N}=1, E_8$ SYM over  \thee  nearly-K{\"a}hler manifold $SU(3)/U(1)\times U(1)$. The $4-$dimensional
gauge group will be derived \by \thee following decomposition of $E_8$ under
$R=U(1)\times U(1)$
\begin{equation}
E_8\supset E_6\times SU(3)\supset E_6\times U(1)_A\times U(1)_B\,.
\end{equation}
Satisfying  \thee  above criterion, \thee surviving
gauge group in four dimensions is
\begin{equation}
H=C_{E_8}(U(1)_A\times U(1)_B)=E_6\times U(1)_A\times U(1)_B\,.
\end{equation}
The surviving scalars \andd fermions in four dimensions are obtained
\by \thee decomposition of \thee adjoint representation of
$E_8$, that is  \thee  $248$, under $U(1)_A\times U(1)_B$. Applying \thee CSDR rules, one obtains  \thee  resulting $4$-dimensional theory, which is an
$\mathcal{N}=1$, $E_6$ GUT, with $U(1)_A, U(1)_B$ global
symmetries. The potential is fully determined after a lengthy calculation and can be found in ref \cite{manousselis-zoupanos2}. Subtracting  \thee  $F-$ and $D-$ terms contributing to this potential, one can determine also scalar masses and trilinear scalar terms, which can be identified with  \thee  scalar part of  \thee  soft supersymmetry breaking sector of  \thee  theory. In addition,  \thee  gaugino obtains a mass, which receives a contribution from  \thee  torsion, contrary to  \thee  rest soft supersymmetry breaking terms. The imortant point to note is that  \thee  CSDR leads to  \thee  soft supersymmetry breaking sector without any additional assumption.

In order to break further  \thee  $E_6$ GUT, one has to employ  \thee  Wilson flux mechanism. Due to  \thee  space limitation we cannot describe here  \thee  mechanism and its application in  \thee  present case. The details can be found in ref \cite{i-z}. The resulting theory is a softly broken $\mathcal{N}=1$, chiral $SU(3)^3$ theory which can break further to an extension of  \thee  MSSM.

\section{Fuzzy spaces}
\subsection{The fuzzy sphere}
In order to introduce  \thee  non-commutative space of \thee fuzzy sphere, we are going to begin with  \thee  familiar ordinary sphere $S^2$ and extend it to its fuzzy version. The $S^2$ may be considered as a manifold embedded into  \thee  $3-$dimensional Euclidean space, $\mathbb{R}^3$. This embedding allows us to specify  \thee  algebra of  \thee  functions on $S^2$ through $\mathbb{R}^3$, \by imposing  \thee  constraint
\begin{equation}
  \sum_{a=1}^3x_a^2=R^2\,,\label{spherecondition}
\end{equation}
where $x_a$ are  \thee  coordinates of $\mathbb{R}^3$ and $R$ is  \thee  radius
of $S^2$. The isometry group of $S^2$ is a global $SO(3)$, which is generated \by  \thee
three angular momentum operators, $L_a=-i\epsilon_{abc}x_b\pa_c$, due to  \thee  isomorphism $SO(3)\simeq SU(2)$.

If we write  \thee  three operators $L_a$ in terms of  \thee  spherical
coordinates $\theta,\phi$, \thee generators are expressed as $L_a=-i\xi_a^\alpha\pa_\alpha$,
where \thee greek index, $\alpha$, denotes \thee spherical coordinates
and $\xi_a^\alpha$ are \thee components of \thee Killing vector fields
which generate \thee isometries of \thee sphere\footnote{The $S^2$
metric can be expressed in terms of \thee Killing vectors as
$g^{\alpha\beta}=\dfrac{1}{R^2}\xi^\alpha_a\xi_a^\beta$.}.

The spherical harmonics, $Y_{lm}(\theta,\phi)$, are \thee eigenfunctions of \thee operator
\begin{equation}
  L^2=-R^2\triangle_{S^2}=-R^2\frac{1}{\sqrt{g}}\pa_a(g^{ab}\sqrt{g}\pa_b)\,.
\end{equation}
Acting with $L^2$ on $Y_{lm}(\theta,\phi)$, one obtains its eigenvalues,
\begin{equation}
  L^2 Y_{lm}=l(l+1)Y_{lm}\,,
\end{equation}
where $l$ is a non-negative integer. In addition, \thee $Y_{lm}(\theta,\phi)$ obey \thee
orthogonality condition
\begin{equation}
  \int \sin\theta d\theta d\phi Y_{lm}^\dag
  Y_{l'm'}=\delta_{ll'}\delta_{mm'}\,.
\end{equation}

Since $Y_{lm}(\theta,\phi)$ form a complete and orthogonal set of functions, any function on $S^2$ can be expanded on this set
\begin{equation}
  a(\theta,\phi)=\sum_{l=0}^\infty\sum_{m=-l}^la_{lm}Y_{lm}(\theta,\phi)\,,\label{sphereexpansion}
\end{equation}
where $a_{lm}$ are complex coefficients. Alternatively, spherical harmonics can also be expressed
in terms of \thee coordinates of $\mathbb{R}^3$, $x_a$, as
\begin{equation}
  Y_{lm}(\theta,\phi)=\sum_{\vec{a}}f^{lm}_{a_1\ldots a_l}x^{a_1\ldots
  a_l}\,,\label{cartesianharmonics}
\end{equation}
where $f^{lm}_{a_1\ldots a_l}$ is an $l-$rank (traceless) symmetric tensor.

Let us now make \thee extension of $S^2$ to its fuzzy version. Fuzzy sphere is a typical case of a non-commutative space, meaning that \thee algebra of functions is not commutative, as it is on $S^2$, with $l$ having an upper limit. Therefore, due to this truncation, one obtains a finite dimensional (non-commutative) algebra, in particular $l^2$ dimensional. Thus, it is natural to consider \thee truncated
algebra as a matrix algebra and it is consistent to consider \thee fuzzy
sphere as a matrix approximation of \thee $S^2$.

According to \thee above, it follows that we are able to expand $N$-dimensional matrices
on a fuzzy sphere as
\begin{equation}
  \hat{a}=\sum_{l=0}^{N-1}\sum_{m=-l}^{l}a_{lm}\hat{Y}_{lm}\,,\label{fuzzyexpansion}
\end{equation}
where $\hat{Y}_{lm}$ are spherical harmonics of \thee fuzzy sphere, which are
now given \by
\begin{equation}
  \hat{Y}_{lm}=R^{-l}\sum_{\vec{a}}f_{a_1\ldots
  a_l}^{lm}\hat{X}^{a_1}\cdots\hat{X}^{a_l}\,,
\end{equation}
where
\begin{equation}
  \hat{X}_a=\frac{2R}{\sqrt{N^2-1}}\lambda_a^{(N)}\,,\label{fuzzycoordinates}
\end{equation}
where $\lambda_a^{(N)}$ are \thee $SU(2)$ generators in \thee
$N$-dimensional representation and $f_{a_1\ldots a_l}^{lm}$ is \thee
same tensor that we met in \eqref{cartesianharmonics}. The $\hat{Y}_{lm}$ also
satisfy \thee orthonormality condition
\begin{equation}
  \text{Tr}_N\left(\hat{Y}_{lm}^\dag\hat{Y}_{l'm'}\right)=\delta_{ll'}\delta_{mm'}\,.
\end{equation}

Moreover, there is a relation between \thee expansion of a function,
\eqref{sphereexpansion}, and that of a matrix,
\eqref{fuzzyexpansion} on \thee original and \thee fuzzy sphere,
respectively
\begin{equation}
  \hat{a}=\sum_{l=0}^{N-1}\sum_{m=-l}^{l}a_{lm}\hat{Y}_{lm}\quad\rightarrow\quad
  a=\sum_{l=0}^{N-1}\sum_{m=-l}^{l}a_{lm}Y_{lm}(\theta,\phi)\,.\label{mapping}
\end{equation}
The above relation is obviously a map from matrices to functions.
Since we introduced \thee fuzzy sphere as a truncation of \thee algebra
of functions on $S^2$, considering \thee same $a_{lm}$ was just \thee
most natural choice. Of course, \thee choice of \thee map is not unique,
since it is not obligatory to consider \thee same expansion
coefficients $a_{lm}$. The above is a $1:1$ mapping given \by\cite{andrews},
\begin{equation}
  a(\theta, \phi)=\sum_{lm}\text{Tr}_N(\hat{Y}_{lm}^\dag\hat{a})Y_{lm}(\theta, \phi)\,.
\end{equation}
The matrix trace is mapped to an integral over \thee sphere:
\begin{equation}
  \frac{1}{N}\text{Tr}_N\quad\rightarrow\quad\frac{1}{4\pi}\int d\Omega\,.
\end{equation}

Summing up, \thee fuzzy sphere is a
matrix approximation of \thee ordinary sphere, $S^2$. The truncation of
\thee algebra of \thee functions results to loss of
commutativity, ending up with a non-commutative algebra, that of
matrices, $\text{Mat}(N;\mathbb{C})$. Therefore, \thee fuzzy sphere,
$S_N$, is \thee non-commutative manifold with $\hat{X}_a$ being \thee
coordinate functions. As given \by \eqref{fuzzycoordinates}, $\hat{X}_a$
are $N\times N$ hermitian matrices produced \by \thee
generators of $SU(2)$ in \thee $N-$dimensional representation. Obviously
they have to obey both \thee condition
\begin{equation}
\sum_{a=1}^3\hat{X}_a\hat{X}_a=R^2\,,\label{fuzzyspherecondition}
\end{equation}
which is \thee analogue of \eqref{spherecondition}, and \thee
commutation relations
\begin{equation}
[\hat{X}_a,\hat{X}_b]=i\alpha\epsilon_{abc}\hat{X}_c\,,\quad
  \alpha=\frac{2R}{\sqrt{N^2-1}}\,.\label{fuzzycommutationrelation}
\end{equation}
Equivalently, one can consider \thee algebra on $S_N$ being
described \by \thee antihermitian matrices
\begin{equation}
  X_a=\frac{\hat{X}_a}{i\alpha R}\,,
\end{equation}
also satisfying \thee modified relations \eqref{fuzzyspherecondition},
\eqref{fuzzycommutationrelation}
\begin{equation}
  \sum_{a=1}^3X_aX_a=-\frac{1}{\alpha^2}\,,\quad
  [X_a,X_b]=C_{abc}X_c\,,
\end{equation}
where $C_{abc}=\dfrac{1}{R}\epsilon_{abc}$\,.

Let us proceed \by briefly mentioning \thee differential calculus on \thee fuzzy
sphere, which is  $3-$dimensional and $SU(2)$ covariant.
The derivations of a function $f$, along $X_a$ are
\begin{equation}
  e_a(f)=[X_a,f]\,,
\end{equation}
and consequently, \thee Lie derivative on $f$ is
\begin{equation}
  \mathcal{L}_af=[X_a,f]\,,\label{liederonfunction}
\end{equation}
where $\mathcal{L}_a$ obeys \thee Leibniz rule and \thee commutation relation of
$\mathfrak{su(2)}$
\begin{equation}
  [\mathcal{L}_a,\mathcal{L}_b]=C_{abc}\mathcal{L}_c\,.\label{liecommutator}
\end{equation}
Working on \thee framework of differential forms, $\theta^a$ are \thee $1-$forms
dual to \thee vector fields $e_a$, namely $\langle e_a,\theta^b\rangle=\delta^b_a$.
Therefore, \thee exterior derivative, $d$, acting on a function $f$, gives
\begin{equation}
  df=[X_a,f]\theta^a\,,
\end{equation}
and \thee action of \thee Lie derivative on \thee $1-$forms
$\theta^b$ gives
\begin{equation}
  \mathcal{L}_a\theta^b=C_{abc}\theta^c\,.\label{liederonform}
\end{equation}
The Lie derivative obeys \thee Leibniz law, therefore its action on any $1-$form $\omega=\omega_a\theta^a$ gives
\begin{equation}
  \mathcal{L}_b\omega=\mathcal{L}_b(\omega_a\theta^a)=[X_b,\omega_a]\theta^a-\omega_aC^a_{bc}\theta^c\,,
\end{equation}
where we have applied \eqref{liederonfunction},
\eqref{liederonform}. Therefore, one obtains \thee result
\begin{equation}
  (\mathcal{L}_b\omega)_a=[X_b,\omega_a]-\omega_cC^c_{ba}\,.
\end{equation}

After having stated \thee differential geometry of fuzzy sphere,
one could extend \thee study of \thee differential geometry of $M^4\times S^2_N$,
which is \thee product of Minkowski space and fuzzy sphere with fuzziness level
$N-1$. For example, any $1-$form $A$ of $M^4\times
S^2_N$ can be expressed in terms of $M^4$ and $S_N^2$, that
is
\begin{equation}
  A=A_\mu dx^\mu+A_a\theta^a\,,\label{gaugepotentialoneform}
\end{equation}
where $A_\mu, A_a$ depend on both $x^\mu$ and $X_a$ coordinates.

Furthermore, instead of functions on \thee fuzzy sphere, one can
examine \thee case of spinors
\cite{aschieri-madore-manousselis-zoupanos}. Moreover, although we do not include them in \thee present review, studies of \thee differential geometry of other higher-dimensional fuzzy spaces (e.g. fuzzy $CP^M$) have been done \cite{aschieri-madore-manousselis-zoupanos}.

\subsection{Gauge theory on \thee fuzzy sphere}

Let us consider a field $\phi(X_a)$ on \thee fuzzy sphere, depending on \thee powers of \thee coordinates, $X_a$. The infinitesimal transformation of $\phi(X_a)$ is
\begin{equation}
  \delta\phi(X)=\lambda(X)\phi(X)\,,\label{gaugetransffuzzy}
\end{equation}
where $\lambda(X)$ is \thee parameter of \thee gauge transformation. If $\lambda(X)$ is an antihermitian function of $X_a$, \thee \eqref{gaugetransffuzzy} is an infinitesimal (abelian) $U(1)$ transformation. On \thee other hand, if $\lambda(X)$ is valued in $\text{Lie}(U(P))$, that is \thee algebra of $P\times P$ hermitian matrices, then \thee \eqref{gaugetransffuzzy} is
infinitesimal (non-abelian), $U(P)$. Naturally, it holds that $\delta X_a=0$, which ensures \thee invariance of \thee covariant derivatives under a gauge
transformation. Therefore, in \thee non-commutative case, left multiplication
\by a coordinate is not a covariant operation, that is
\begin{equation}
  \delta(X_a\phi)=X_a\lambda(X)\phi\,,
\end{equation}
and in general it holds that
\begin{equation}
  X_a\lambda(X)\phi\neq\lambda(X)X_a\phi\,.
\end{equation}
Motivated \by \thee non-fuzzy gauge theory, one may introduce \thee
covariant coordinates $\phi_a$, such that
\begin{equation}
  \delta(\phi_a\phi)=\lambda\phi_a\phi\,,
\end{equation}
which holds if
\begin{equation}
  \delta(\phi_a)=[\lambda,\phi_a]\,.\label{transformationofphia}
\end{equation}
Usual (non-fuzzy) gauge theory also
guides one to define
\begin{equation}
  \phi_a\equiv X_a+A_a\,,\label{covariantfield}
\end{equation}
with \thee $A_a$ being interpreted as \thee gauge potential of \thee
non-commutative theory. Therefore, \thee covariant coordinates $\phi_a$ are \thee non-commutative analogue of \thee covariant derivative of ordinary
gauge theories. From \eqref{covariantfield},
\eqref{transformationofphia} one is led to \thee transformation of
$A_a$, that is
\begin{equation}
  \delta A_a=-[X_a,\lambda]+[\lambda,A_a]\,,
\end{equation}
a form that encourages \thee interpretation of $A_a$ as a gauge field.
In correspondence with \thee non-fuzzy gauge theory, one proceeds with defining a field strength tensor, $F_{ab}$, as
\begin{equation}
  F_{ab}\equiv[X_a,A_b]-[X_b,A_a]+[A_a,A_b]-C^c_{ab}A_c=[\phi_a,\phi_b]-C^c_{ab}\phi_c
\end{equation}
It can be proven that \thee transformation of \thee above field
strength tensor is covariant:
\begin{equation}
  \delta F_{ab}=[\lambda,F_{ab}]\,.
\end{equation}
\section{Ordinary fuzzy dimensional reduction and gauge symmetry enhancement}

Let us now proceed \by performing a simple (trivial) dimensional reduction in order to demonstrate \thee structure we sketched in \thee previous section. Starting with a higher-dimensional theory on $M^4\times (S/R)_F$, with gauge group $G=U(P)$, we determine \thee produced $4-$dimensional theory after performing \thee reduction and finally we make comments on \thee results. Let
$(S/R)_F$ be a fuzzy coset, e.g. \thee fuzzy sphere, $S^2_N$. The
action is
\begin{equation}
  \mathcal{S}_{YM}=\frac{1}{4g^2}\int
  d^4xk\text{Tr}\text{tr}_GF_{MN}F^{MN}\,,\label{actiondimred}
\end{equation}
with $\text{tr}_G$ \thee trace of \thee gauge group $G$ and
$k\text{Tr}$\footnote{In general, $k$ is a parameter related to \thee
size of \thee fuzzy coset space. In \thee case of \thee fuzzy
sphere, $k$ is related to \thee radius of \thee
sphere and \thee integer $l$.} denotes \thee integration over $(S/R)_F$,
i.e. \thee fuzzy coset which is described \by $N\times N$ matrices.
$F_{MN}$ is \thee higher-dimensional field strength tensor, which
is composed of both $4-$dimensional spacetime and extra-dimensional
parts, i.e. $(F_{\mu\nu},F_{\mu a},F_{ab})$.
The components of $F_{MN}$ in \thee extra
(non-commutative) directions, are expressed in terms of \thee covariant
coordinates $\phi_a$, as follows
\begin{align}
  F_{\mu a}&=\pa_\mu\phi_a+[A_\mu,\phi_a]=D_\mu\phi_a\nonumber\\
  F_{ab}&=[X_a,A_b]-[X_b,A_a]+[A_a,A_b]-C_{ba}^cA_{ac}\,.\nonumber
\end{align}
Putting \thee above equations in \eqref{actiondimred}, \thee action
takes \thee form
\begin{equation}
  \mathcal{S}_{YM}=\int
  d^4x\text{Tr}\text{tr}_G\left(\frac{k}{4g^2}F_{\mu\nu}^2+\frac{k}{2g^2}(D_\mu\phi_a)^2-V(\phi)\right)\,,\label{action2}
\end{equation}
where $V(\phi)$ denotes \thee potential, derived from \thee kinetic
term of $F_{ab}$, that is
\begin{align}
  V(\phi)&=-\frac{k}{4g^2}\text{Tr}\text{tr}_G\sum_{ab}F_{ab}F_{ab}\nonumber\\
  &=-\frac{k}{4g^2}\text{Tr}\text{tr}_G\left([\phi_a,\phi_b][\phi^a,\phi^b]-4C_{abc}\phi^a\phi^b\phi^c+2R^{-2}\phi^2\right)\,.
\end{align}
It is natural to consider \eqref{action2} as an action of a $4-$dimensional theory. Let $\lambda(x^\mu,X^a)$ be \thee gauge parameter that appears in an infinitesimal gauge transformation of $G$. This transformation can be interpreted as a $M^4$ gauge transformation. We write
\begin{equation}
  \lambda(x^\mu,X^a)=\lambda^I(x^\mu,X^a)\mathcal{T}^I=\lambda^{h,I}(x^\mu)\mathcal{T}^h\mathcal{T}^I\,,\label{lambdareduction}
\end{equation}
where $\mathcal{T}^I$ denote \thee hermitian generators of \thee gauge
group $U(P)$. $\lambda^I(x^\mu,X^a)$ are \thee $N\times N$ antihermitian
matrices, therefore they can be expressed as
$\lambda^{I,h}(x^\mu)\mathcal{T}^h$, where $\mathcal{T}^h$ are \thee antihermitian
generators of $U(N)$ and $\lambda^{I,h}(x^\mu), h=1,\ldots, N^2$,
are \thee Kaluza-Klein modes of $\lambda^I(x^\mu,X^a)$. In turn, we can assume that \thee fields on \thee right hand side of
\eqref{lambdareduction} could be considered as one field that takes
values in \thee tensor product Lie algebra
$\text{Lie}\left(U(N)\right)\otimes\text{Lie}\left(U(P)\right)$,
which corresponds to \thee algebra
$\text{Lie}\left(U(NP)\right)$. Similarly, \thee gauge field $A_\nu$ can
be written as
\begin{equation}
  A_\nu(x^\mu,X^a)=A_\nu^I(x^\mu,X^a)\mathcal{T}^I=A_\nu^{h,I}(x^\mu)\mathcal{T}^h\mathcal{T}^I\,,
\end{equation}
which is interpreted as a gauge field on $M^4$ that takes values in \thee $\text{Lie}\left(U(NP)\right)$ algebra. A similar
consideration can also be applied in \thee case of scalar fields \footnote{Also, $\text{Tr}\text{tr}_G$ is interpreted as
\thee trace of \thee $U(NP)$ matrices.}.

It is worth noting \thee enhancement of \thee gauge symmetry of \thee $4-$dimensional theory compared to \thee gauge symmetry of \thee higher-dimensional theory. In other words, we can start with an abelian gauge group in higher dimensions and result with a non-abelian gauge symmetry in \thee $4-$dimensional theory. A defect of this theory is that \thee scalars are accommodated in \thee adjoint representation of \thee $4-$dimensional gauge group, which means that they cannot induce \thee electroweak symmetry breaking. This motivates \thee realization of non-trivial
dimensional reduction schemes, like \thee one that follows in \thee next section.

\section{Fuzzy CSDR}

In order to result with a less defective $4-$dimensional gauge theory, we proceed \by performing a non-trivial dimensional reduction, that is \thee fuzzy version of \thee CSDR.

So, in this section we adopt \thee CSDR programme in \thee non-commutative framework, where \thee extra dimensions are fuzzy coset spaces
\cite{aschieri-madore-manousselis-zoupanos}\footnote{See also
\cite{Harland-kurkcuoglu}.}, in order to result with a smaller number of both gauge and scalar fields in \thee $4-$dimensional action
\eqref{action2}. In general, \thee group $S$ acts on \thee fuzzy coset
$(S/R)_F$, and in accordance with \thee commutative case, CSDR scheme
suggests that \thee fields of \thee theory must be invariant under an
infinitesimal group $S-$transformation, up to an infinitesimal
gauge transformation. Specifically, \thee fuzzy coset in this case is \thee
fuzzy sphere, $(SU(2)/U(1))_F$, so \thee action of an infinitesimal
$SU(2)-$transformation should leave \thee scalar
and gauge fields invariant, up to an infinitesimal gauge
transformation
\begin{align}
  \mathcal{L}_b\phi&=\delta_{W_b}=W_b\phi \label{csdrconstraintone} \\
  \mathcal{L}_bA&=\delta_{W_b}A=-DW_b\,, \label{csdrconstrainttwo}
\end{align}
where $A$ is \thee gauge potential expressed as an $1-$form, see
\eqref{gaugepotentialoneform}, and $W_b$ is an antihermitian gauge parameter depending only on \thee coset coordinates $X^a$.
Therefore, $W_b$ is written as
\begin{equation}
  W_b=W_b^I\mathcal{T}^I\,,\quad I=1,2,\ldots, P^2\,,
\end{equation}
where $\mathcal{T}^I$ are \thee hermitian generators of $U(P)$ and
$(W_b^I)^\dag=-W_b^I$, where \thee $^\dag$ denotes \thee hermitian
conjugation on \thee $X^a$ coordinates.

Putting into use \thee covariant coordinates, $\phi_a$, \eqref{covariantfield},
and $\omega_a$, defined as
\begin{equation}
  \omega_a\equiv X_a-W_a\,,
\end{equation}
\thee CSDR constraints, \eqref{csdrconstraintone},
\eqref{csdrconstrainttwo}, convert to
\begin{align}
  [\omega_b,A_\mu]&=0\label{constraintone}\\
  C_{bde}\phi^e&=[\omega_b,\phi_d]\label{constrainedtwo}\,.
\end{align}
Due to \thee fact that Lie derivatives respect \thee $\mathfrak{su(2)}$ commutation relation, \eqref{liecommutator}, one results with \thee following consistency condition
\begin{equation}
  [\omega_a,\omega_b]=C_{ab}^c\omega_c\,,\label{consistencycondition}
\end{equation}
where \thee transformation of $\omega_a$ is given \by
\begin{equation}
  \omega_a\,\rightarrow\,\omega_a'=g\omega_ag^{-1}\,.
\end{equation}
In \thee case of spinor fields, \thee procedure is quite similar
\cite{aschieri-madore-manousselis-zoupanos}.

Let us now consider a higher-dimensional theory with gauge symmetry $G=U(1)$. We are going to perform a fuzzy CSDR, in which \thee fuzzy sphere is $(S/R)_F=S_N^2$. The $\omega_a=\omega_a(X^b)$ are
$N\times N$ antihermitian matrices, therefore they can be
considered as elements of $\text{Lie}(U(N))$. At \thee same time, they satisfy \thee commutation relation of $\text{Lie}(SU(2))$, as in \thee consistency condition, \eqref{consistencycondition}. So we have to embed $\text{Lie}(SU(2))$ into $\text{Lie}(U(N))$. Therefore, if $\mathcal{T}^h,\, h=1,\ldots,N^2$ are \thee $\text{Lie}(U(N))$
generators, in \thee fundamental representation, then \thee convention $h=(a,u),\,a=1,2,3\,,\,u=4,5,\ldots,N^2$ can be used, obviously with
\thee generators $T^a$ satisfying $\text{Lie}(SU(2))$
\begin{equation}
  [T^a,T^b]=C^{ab}_cT^c\,.\label{relationtorespect}
\end{equation}
At last, \thee embedding is defined \by \thee identification
\begin{equation}
  \omega_a=T_a\,.
\end{equation}
So, \thee constraint \eqref{constraintone} implies that \thee gauge
group of \thee $4-$dimensional theory, $K$, is \thee centralizer of \thee
image of $SU(2)$ into $U(N)$, that is
\begin{equation}
  K=C_{U(N)}(SU(2))=SU(N-2)\times U(1)\times U(1)\,,
\end{equation}
where \thee second $U(1)$ in \thee right hand side is present due to
\begin{equation}
  U(N)\simeq SU(N)\times U(1)\,.
\end{equation}
Therefore, $A_\mu(x,X)$ are arbitrary functions over $x$, but
they depend on $X$ in a way that take values in
$\text{Lie}(K)$ instead of $\text{Lie}(U(N))$. That means
that we result with a $4-$dimensional gauge potential which takes values in $\text{Lie}(K)$.

Let us now study \thee next constraint, \eqref{constrainedtwo}.
This one gets satisfied choosing
\begin{equation}
  \phi_a=r\phi(x)\omega_a\,,
\end{equation}
meaning that \thee degrees of freedom remaining unconstrained
are related to \thee scalar field, $\phi(x)$, which is singlet under \thee $4-$dimensional gauge group, $K$.

Summing up \thee results from \thee above reduction, \thee consistency condition \eqref{consistencycondition}, dictated \thee embedding of $SU(2)$ into $U(N)$. Although \thee embedding was realized into \thee fundamental representation of $U(N)$, we could have used \thee irreducible $N$-dimensional representation of $SU(2)$ \by identifying
$\omega_a=X_a$. If so, \thee constraint \eqref{constraintone}
would lead to \thee $U(1)$ to be \thee $4-$dimensional gauge group,
with $A_\mu(x)$ getting values in $U(1)$. The second constraint,
\eqref{constrainedtwo}, implies that, in this case too, $\phi(x)$ is
a scalar singlet.

To conclude \thee whole procedure, one starts with a $U(1)$ higher-dimensional gauge theory on $M^4\times S^2_N$ and because of \thee consistency condition, \eqref{consistencycondition}, an embedding of $SU(2)$ into $U(N)$ is required\footnote{This embedding is achieved non-uniquely, specifically in $p_N$ ways, where $p_N$ is \thee possible ways one can partition \thee $N$ into a set of non-increasing, positive
integers \cite{madore}.}. So, \thee first fuzzy CSDR constraint,
\eqref{constraintone}, gives \thee $4-$dimensional gauge group and
from \thee second one, \eqref{constrainedtwo}, one obtains \thee
$4-$dimensional scalar fields, surviving from \thee dimensional reduction.

As far as \thee fermions are concerned, we briefly mention
\thee results of \thee above dimensional reduction. According to \thee extended analysis \cite{aschieri-madore-manousselis-zoupanos}, it is proven that \thee appropriate choice of embedding is
\begin{equation}
  S\subset SO(\text{dim}S)\,,
\end{equation}
which is achieved \by $T_a=\dfrac{1}{2}C_{abc}\Gamma^{bc}$, respecting \eqref{relationtorespect}. Therefore, $\psi$ functions as an interwining operator between \thee representations of $S$ and $SO(\text{dim}S)$.
In accordance to \thee commutative (non-fuzzy) case, \cite{kapetanakis-zoupanos}, in order to find \thee surviving fermions in \thee $4-$dimensional theory, one has to decompose \thee adjoint representation of $U(NP)$ under \thee product $S_{U(NP)}\times K$, that is
\begin{align}
  U(NP)\supset &\,S_{U(NP)}\times K\,,\\
 \text{adj}\,U(NP)&=\sum_i(s_i,k_i)\,.
\end{align}
Also, \thee decomposition of \thee spinorial representation $\sigma$ of
$SO(\text{dim}S)$ under $S$ is
\begin{align}
  SO&(\text{dim}S)\supset S\,,\\
  \sigma&=\sum_e\sigma_e\,.
\end{align}
Thus, if \thee two irreducible representations
$s_i, \sigma_e$ are identical, \thee surviving fermions
of \thee $4-$dimensional theory ($4-$dimensional spinors) belong to
\thee $k_i$ representation of gauge group $K$.

Before we move on, this is a suitable point to compare \thee higher-dimensional theory $M^4\times (S/R)$, to its fuzzy extension, $M^4\times (S/R)_F$. The first similarity has to do with \thee fact that fuzziness does not affect \thee isometries, both spaces have \thee same, $SO(1,3)\times SO(3)$. The second is that \thee gauge couplings defined on both spaces have \thee same dimensionality. But, on \thee other hand, a very striking difference is that of \thee two, only \thee non-commutative higher-dimensional theory is renormalizable \footnote{The number of counter-terms required to eliminate \thee divergencies is finite.}. In addition, a $U(1)$ initial gauge symmetry on $M^4\times(S/R)_F$, is enough in order to result with non-abelian
structures in four dimensions\footnote{Technically, this is possible
because $N\times N$ matrices can be decomposed on \thee $U(N)$
generators.}.

\section{Orbifolds and fuzzy extra dimensions}

The introduction of \thee orbifold structure (similar to \thee one developed in
\cite{kachru-silverstein}) in \thee framework of gauge theories with fuzzy extra dimensions was motivated \by \thee necessity of chiral low energy theories. In order to justify further \thee renormalizability of \thee theories constructed so far using fuzzy extra dimensions, we were led to consider \thee reverse procedure and start from a renormalizable theory in four dimensions and try to reproduce \thee results of a higher-dimensional theory reduced over fuzzy coset spaces \cite{aschieri-grammatikopoulos, steinacker-zoupanos,
chatzistavrakidis-steinacker-zoupanos}. This idea was realized as follows: one starts with a $4-$dimensional gauge theory including appropriate scalar fields and a suitable potential leading to vacua that could be interpreted as dynamically generated fuzzy extra dimensions, including a finite Kaluza-Klein tower of massive modes. This reverse procedure gives hope that an initial abelian gauge theory does not have to be higher-dimensional and \thee non-abelian gauge theory structure could emerge from fluctuations of \thee coordinates \cite{steinacker1}. The whole idea eventually seems to have similarities to \thee idea of dimensional deconstruction introduced earlier \cite{kim}.

\begin{comment}
The concept of deconstructing dimensions \cite{arkani},
motivated \thee idea to reverse \thee above
\cite{aschieri-grammatikopoulos, steinacker-zoupanos,
chatzistavrakidis-steinacker-zoupanos} in order to further justify
\thee renormalizability of \thee theory but also to attempt \thee construction of
chiral models in theories arising from \thee framework of fuzzy extra
dimensions. Moreover, \thee reversed procedure gives hope that \thee initial abelian gauge theory does not have to be higher-dimensional with \thee non-abelian gauge theory emerging from fluctuations of \thee coordinates \cite{steinacker1}. This
idea is realized as follows: one begins with a
$4-$dimensional gauge theory including an appropriate scalar
spectrum and a suitable potential leading to vacua that could
be interpreted as dynamically generated fuzzy extra dimensions also including
a finite Kaluza-Klein tower of massive modes.
\end{comment}
The inclusion of fermions in such models was desired too, but \thee
best one could achieve for some time contained mirror fermions in bifundamental representations of \thee low-energy gauge group \cite{steinacker-zoupanos,chatzistavrakidis-steinacker-zoupanos}.
Mirror fermions do not exclude \thee possibility to make contact with
phenomenology \cite{maalampi-roos}, nevertheless, it is preferred to
result with exactly chiral fermions.

Next, \thee plan that was sketched above is realized. Specifically, we are going to deal with \thee $\mathbb{Z}_3$ orbifold projection of \thee $\mathcal{N}=4$ Supersymmetric Yang Mills (SYM)
theory \cite{brink-schwarz-scherk}, examining \thee action of \thee
discrete group on \thee fields of \thee theory and \thee superpotential that
emerges in \thee projected theory.

\subsection{$\mathcal{N}=4$ SYM field theory and $\mathbb{Z}_3$ orbifolds \label{subsec:subsection}}

So, let us begin with an $\mathcal{N}=4$ supersymmetric $SU(3N)$
gauge theory defined on \thee Minkowski spacetime. The particle spectrum of \thee theory (in \thee $\mathcal{N}=1$ terminology) consists of an $SU(3N)$ gauge
supermultiplet and three adjoint chiral supermultiplets
$\Phi^i\,,i=1,2,3$. The component fields of \thee above
supermultiplets are \thee gauge bosons, $A_\mu,\,\mu=1,\ldots,4$, six
adjoint real (or three complex) scalars $\phi^a,\,a=1,\ldots,6$ and
four adjoint Weyl fermions $\psi^p,\,p=1,\ldots,4$. The scalars and
Weyl fermions transform according to \thee $6$ and $4$ representations of \thee
$SU(4)_R$ $R$-symmetry of \thee theory, respectively, while \thee gauge bosons are singlets.

Then, in order to introduce orbifolds, \thee discrete group
$\mathbb{Z}_3$ has to be considered as a subgroup of $SU(4)_R$. The choice of \thee
embedding of $\mathbb{Z}_3$ into $SU(4)_R$ is not unique and \thee
options are not equivalent, since \thee choice of embedding
affects \thee amount of \thee remnant supersymmetry
\cite{kachru-silverstein}:
\begin{itemize}
  \item Maximal embedding of $\mathbb{Z}_3$ into $SU(4)_R$ is excluded because it
  would lead to non-supersymmetric models,
  \item Embedding of $\mathbb{Z}_3$ in an $SU(4)_R$ subgroup:
  \begin{itemize}
    \item[-] Embedding into an $SU(2)$ subgroup would lead to
    $\mathcal{N}=2$ supersymmetric models with $SU(2)_R$ $R$-symmetry
    \item[-] Embedding into an $SU(3)$ subgroup would lead to
    $\mathcal{N}=1$ supersymmetric models with $U(1)_R$
    $R$-symmetry.
  \end{itemize}
\end{itemize}

We focus on \thee last embedding which is \thee desired one, since it leads to
$\mathcal{N}=1$ supersymmetric models. Let us consider a generator
$g\in\mathbb{Z}_3$, labeled (for convenience) \by three
integers $\vec{a}=(a_1, a_2, a_3)$ \cite{douglas-greene-morrison}
satisfying \thee relation
\begin{equation}
  a_1+a_2+a_3=0\,\,\text{mod}\,3\,.
\end{equation}
The last equation implies that $\mathbb{Z}_3$ is embedded
in \thee $SU(3)$ subgroup, i.e. \thee remnant supersymmetry is \thee desired
$\mathcal{N}=1$ \cite{bailin-love}.

It is expected that since \thee various fields of \thee theory transform
differently under $SU(4)_R$, $\mathbb{Z}_3$ will act
non-trivially on them. Gauge and gaugino fields are singlets under $SU(4)_R$,
therefore \thee geometric action of \thee $\mathbb{Z}_3$ rotation is
trivial. The action of $\mathbb{Z}_3$ on \thee complex scalar fields is
given \by \thee matrix $\gamma(g)_{ij}=\delta_{ij}\omega^{a_i}$,
where $\omega=e^{\frac{2\pi}{3}}$ and \thee action of $\mathbb{Z}_3$ on \thee fermions $\phi^i$ is
given \by $\gamma(g)_{ij}=\delta_{ij}\omega^{b_i}$, where $b_i=-\dfrac{1}{2}(a_{i+1}+a_{i+2}-a_i)$\footnote{Also modulo 3}. In \thee case under study \thee three integers of \thee generator $g$ are
$(1,1,-2)$, meaning that $a_i=b_i$.

The matter fields are not invariant under a gauge transformation, therefore $\mathbb{Z}_3$ acts on their gauge indices, too. The action of this rotation is given \by \thee matrix
\begin{equation}
  \gamma_3=\left(
             \begin{array}{ccc}
               \mathbf{1}_N & 0 & 0 \\
               0 & \omega\mathbf{1}_N & 0 \\
               0 & 0 & \omega^2\mathbf{1}_N \\
             \end{array}
           \right)\,.\label{gammatria}
\end{equation}

There is no specific reason for these blocks to have \thee same
dimensionality (see e.g.\cite{aldabaz-ibanez,
lawrence-nekrason-vafa, kiritsis}). However, since \thee projected
theory must be free of anomalies, \thee dimension of \thee three blocks is
\thee same.

After \thee orbifold projection, \thee spectrum of \thee theory consists of \thee
fields that are invariant under \thee combined action of \thee discrete
group, $\mathbb{Z}_3$, on \thee "geometric"\footnote{In case of
ordinary reduction of a $10$-dimensional $\mathcal{N}=1$
SYM theory, one obtains an $\mathcal{N}=4$
SYM Yang-Mills theory in four dimensions having a
global $SU(4)_R$ symmetry which is identified with \thee tangent space
$SO(6)$ of \thee extra dimensions \cite{manousselis-zoupanos,
manousselis-zoupanos2}.} and gauge indices
\cite{douglas-greene-morrison}. As far as \thee gauge bosons are
concerned being singlets, \thee projection is $A_\mu=\gamma_3 A_\mu\gamma_3^{-1}$.
Therefore, taking into consideration \eqref{gammatria}, \thee gauge group of
\thee initial theory breaks down to \thee group $H=SU(N)\times
SU(N)\times SU(N)$ in \thee projected theory.

As we have already stated, \thee complex scalar fields transform
non-trivially under \thee gauge and $R-$symmetry, so \thee
projection is $\phi_{IJ}^i=\omega^{I-J+a_i}\phi^i_{IJ}$,
where $I,J$ are gauge indices. Therefore, $J=I+a_i$, meaning
that \thee scalar fields surviving \thee orbifold projection have \thee
form $\phi_{I,J+a_i}$ and transform under \thee gauge group
$H$ as
\begin{equation}
  3\cdot\left((N,\bar{N},1)+(\bar{N},1,N)+(1,N,\bar{N})\right)\,.\label{repaftrprojection}
\end{equation}

Similarly, fermions transform non-trivially under \thee gauge group
and $R-$symmetry, too with \thee projection being $\psi^i_{IJ}=\omega^{I-J+b_i}\psi_{IJ}^i$.
Therefore, \thee fermions surviving \thee projection have
\thee form $\psi^i_{I,I+b_i}$ accommodated in \thee same
representation of $H$ as \thee scalars, that is
\eqref{repaftrprojection}, a fact demonstrating \thee $\mathcal{N}=1$ remnant
supersymmetry. It is worth noting that \thee representations \eqref{repaftrprojection} of \thee
resulting theory are anomaly free.

The fermions, summing up \thee above results, are accommodated into
chiral representations of $H$ and there are three fermionic
generations since, as we have mentioned above, \thee particle spectrum
contains three $\mathcal{N}=1$ chiral supermultiplets.

The interactions of \thee projected model are given \by \thee superpotential.
In order to specify it, one has to begin with \thee superpotential of \thee initial
$\mathcal{N}=4$ SYM theory \cite{brink-schwarz-scherk}
\begin{equation}
  W_{\mathcal{N}=4}=\epsilon_{ijk}\text{Tr}(\Phi^i\Phi^j\Phi^k)\,,
\end{equation}
where, $\Phi^i,\Phi^j,\Phi^k$ are \thee three chiral superfields
of \thee theory. After \thee projection, \thee structure of \thee
superpotential remains unchanged, but it encrypts only \thee
interactions among \thee surviving fields of \thee $\mathcal{N}=1$
theory, that is
\begin{equation}
  W_{\mathcal{N}=1}^{(proj)}=\sum_I\epsilon_{ijk}\Phi_{I,I+a_i}^i\Phi_{I+a_i,I+a_i+a_j}^j\Phi_{I+a_i+a_j,I}^k\,.\label{projectedsuper}
\end{equation}

\subsection{Dynamical generation of twisted fuzzy
spheres}

From \thee superpotential $W_{\mathcal{N}=1}^{proj}$ that is given in \eqref{projectedsuper},
\thee scalar potential can be extracted:
\begin{equation}
  V_{\mathcal{N}=1}^{proj}(\phi)=\frac{1}{4}\text{Tr}\left([\phi^i,\phi^j]^\dag[\phi^i,\phi^j]\right)\,,\label{scalarpotential}
\end{equation}
where, $\phi^i$ are \thee scalar component fields of \thee
superfield $\Phi^i$. The potential $V_{\mathcal{N}=1}^{proj}(\phi)$ is minimized \by vanishing
vevs of \thee fields, so modifications have to be made, in order that solutions interpreted as
vacua of a non-commutative geometry to be emerged.

So, in order to result with a minimum of $V_{\mathcal{N}=1}^{proj}(\phi)$,
%which
%is compatible to \thee twisted fuzzy sphere relations,
%\eqref{twistedfuzzy}, we modify \thee theory, firstly \by
$\mathcal{N}=1$ soft supersymmetric terms of \thee form\footnote{The
SSB terms that will be inserted into $V_{\mathcal{N}=1}^{proj}(\phi)$, are
purely scalar. Although this is enough for our purpose, it is obvious
that more SSB terms have to be included too, in order to obtain \thee
full SSB sector \cite{djouadi}.}
\begin{equation}
  V_{SSB}=\frac{1}{2}\sum_im_i^2\phi^{i\dag}\phi^i+\frac{1}{2}\sum_{i,j,k}h_{ijk}\phi^i\phi^j\phi^k+h.c.\label{ssbterms}
\end{equation}
are introduced, where $h_{ijk}=0$ unless
$i+j+k\equiv0\,\text{mod}3$. The introduction of these SSB terms
should not come as a surprise, since \thee presence of an SSB sector is necessary
anyway for a model with realistic aspirations, see
e.g.\cite{djouadi}. The inclusion of \thee $D$-terms of
\thee theory is necessary and they are given \by
\begin{equation}
  V_D=\frac{1}{2}D^2=\frac{1}{2}D^ID_I\,,
\end{equation}
where $D^I=\phi_i^\dag T^I\phi^i$, where $T^I$ are \thee generators in
\thee representation of \thee corresponding chiral multiplets.

So, \thee total potential of \thee theory is given \by
\begin{equation}
V=V_{\mathcal{N}=1}^{proj}+V_{SSB}+V_D\,.\label{totalscalarpotential}
\end{equation}
%Since \thee aim is to obtain twisted fuzzy sphere vacua,
A suitable choice for \thee parameters $m_i^2$ and $h_{ijk}$ in
\eqref{ssbterms} is $m_i^2=1,\,\, h_{ijk}=\epsilon_{ijk}$.
Therefore, \thee total scalar potential, \eqref{totalscalarpotential},
takes \thee form
\begin{equation}
  V=\frac{1}{4}(F^{ij})^\dag F^{ij}+V_D\,,\label{twistedpotential}
\end{equation}
where $F^{ij}$ is defined as
\begin{equation}
  F^{ij}=[\phi^i,\phi^j]-i\epsilon^{ijk}(\phi^k)^\dag\,.\label{fuzzyfieldstrength}
\end{equation}

The first term of \thee scalar potential, \eqref{twistedpotential}, is always positive, therefore, \thee global
minimum of \thee potential is obtained when
\begin{equation}
  [\phi^i,\phi^j]=i\epsilon_{ijk}(\phi^k)^\dag\,,\quad
  \phi^i(\phi^j)^\dag=R^2\,,\label{twistedfuzzy}
\end{equation}
where $(\phi^i)^\dag$ denotes \thee hermitian conjugate of $\phi^i$ and $[R^2,\phi^i]=0$. It is clear that \thee
above equations are related to a fuzzy sphere. This becomes more transparent \by
considering \thee untwisted fields $\tilde{\phi}^i$, defined \by
\begin{equation}
  \phi^i=\Omega\tilde{\phi}^i\,,\label{twisted-untwisted}
\end{equation}
where $\Omega\neq1$ satisfy \thee relations
\begin{align}
\Omega^3=1\,,\,\,\,\,[\Omega,\phi^i]=0\,,\,\,\,\,\Omega^\dag=\Omega^{-1}\,,\,\,\,\,
(\tilde{\phi}^i)^\dag=\tilde{\phi}^i\,\,\Leftrightarrow\,\,
(\phi^i)^\dag=\Omega\phi^i\,.
\end{align}
Therefore, \eqref{twistedfuzzy} reproduces \thee ordinary fuzzy sphere
relations generated \by $\tilde{\phi}^i$
\begin{equation}
  [\tilde{\phi}^i,\tilde{\phi}^j]=i\epsilon_{ijk}\tilde{\phi}^k\,,\quad\tilde{\phi}^i\tilde{\phi}^i=R^2\,,\label{untwisted}
\end{equation}
exhibiting \thee reason why \thee non-commutative space generated \by
$\phi^i$ is a twisted fuzzy sphere, $\tilde{S}_N^2$.

Next, one can find configurations of \thee twisted
fields $\phi^i$, i.e. fields satisfying \eqref{twistedfuzzy}.
Such configuration is
\begin{equation}
  \phi^i=\Omega(\mathbf{1}_3\otimes\lambda^i_{(N)})\,,
\end{equation}
where $\lambda^i_{(N)}$ are \thee $SU(2)$ generators in \thee
$N$-dimensional irreducible representation and $\Omega$ is \thee
matrix
\begin{equation}
  \Omega=\Omega_3\otimes\mathbf{1}_N\,,\quad\Omega_3=\left(
                                                       \begin{array}{ccc}
                                                         0 & 1 & 0 \\
                                                         0 & 0 & 1 \\
                                                         1 & 0 & 0 \\
                                                       \end{array}
                                                     \right)\,,\quad\Omega^3=\mathbf{1}\,.
\end{equation}
According to \thee transformation \eqref{twisted-untwisted}, \thee
"off-diagonal" orbifold sectors \eqref{repaftrprojection} convert to \thee
block-diagonal form
\begin{equation}
  \phi^i=\left(
           \begin{array}{ccc}
             0 & (\lambda_{(N)}^i)_{(N,\bar{N},1)} & 0 \\
             0 & 0 & (\lambda_{(N)}^i)_{(1,N,\bar{N})} \\
             (\lambda_{(N)}^i)_{(\bar{N},1,N)} & 0 & 0 \\
           \end{array}
         \right)=\Omega\left(
                         \begin{array}{ccc}
                           \lambda^i_{(N)} & 0 & 0 \\
                           0 & \lambda^i_{(N)} & 0 \\
                           0 & 0 & \lambda^i_{(N)} \\
                         \end{array}
                       \right)\,.\label{triaseena}
\end{equation}
Therefore, \thee untwisted fields generating \thee
ordinary fuzzy sphere, $\tilde{\phi}^i$, are written in a
block-diagonal form. Each block can be considered as an ordinary fuzzy
sphere, since they separately satisfy \thee corresponding commutation
relations \eqref{untwisted}. In turn, \thee above configuration in
\eqref{triaseena}, which corresponds to \thee vacuum of \thee theory, has \thee form of three fuzzy spheres, appearing with relative angles $2\pi/3$. Concluding, \thee
solution $\phi^i$ can be considered as \thee twisted equivalent of three
fuzzy spheres, conforming with \thee orbifolding.

Note that \thee $F^{ij}$ defined in \eqref{fuzzyfieldstrength}, can
be interpreted as \thee field strength of \thee spontaneously generated
fuzzy extra dimensions. The second term of \thee potential, $V_D$,
induces a change on \thee radius of \thee sphere (in a similar way to \thee case of \thee
ordinary fuzzy sphere \cite{aschieri-grammatikopoulos, steinacker,
chatzistavrakidis-steinacker-zoupanos}).

\subsection{Chiral models after \thee fuzzy orbifold projection - The $SU(3)_c\times SU(3)_L\times SU(3)_R$ model}

The resulting unification groups after \thee orbifold projection are various
because of \thee different ways \thee gauge group $SU(3N)$ is spontaneously broken. The minimal, anomaly free models are $SU(4)\times SU(2)\times SU(2)$,
$SU(4)^3$ and $SU(3)^3$\footnote{Similar approaches have been
studied in \thee framework of YM matrix models \cite{grosse-lizzi},
lacking phenomenological viability.}.

We focus on \thee breaking of \thee latter,
which is \thee trinification group $SU(3)_c\times SU(3)_L\times
SU(3)_R$ \cite{glashow, rizov} (see also
\cite{ma-mondragon-zoupanos, lazarides-panagiotakopoulos, MaZoup,
babu-he-pakvasa, leontaris-rizos} and for a string theory approach
see \cite{kim}). At first, \thee integer $N$ has to be decomposed as
$N=n+3$. Then, for $SU(N)$, \thee considered embedding is
\begin{equation}
  SU(N)\supset SU(n)\times SU(3)\times U(1)\,,\label{decomp}
\end{equation}
from which it follows that \thee embedding for \thee
gauge group $SU(N)^3$ is
\begin{equation}
  SU(N)^3\supset SU(n)\times SU(3)\times SU(n)\times SU(3)\times
  SU(n)\times SU(3)\times U(1)^3\,.\label{decomposition}
\end{equation}
The three $U(1)$ factors are ignored\footnote{As anomalous gaining
mass \by \thee Green-Schwarz mechanism and therefore they decouple at
\thee low energy sector of \thee theory \cite{lawrence-nekrason-vafa}.} and \thee representations are decomposed according to \eqref{decomposition}, (after
reordering \thee factors) as
\begin{align}
  &SU(n)\times SU(n)\times SU(n)\times SU(3)\times SU(3)\times
  SU(3)\,,\nonumber\\
 &(n,\bar{n},1;1,1,1)+(1,n,\bar{n};1,1,1)+(\bar{n},1,n;1,1,1)+(1,1,1;3,\bar{3},1)\nonumber\\
 &+(1,1,1;1,3,\bar{3})+(1,1,1;\bar{3},1,3)+(n,1,1;1,\bar{3},1)+(1,n,1;1,1,\bar{3})\nonumber\\
 &+(1,1,n;\bar{3},1,1)+(\bar{n},1,1;1,1,3)+(1,\bar{n},1;3,1,1)+(1,1,\bar{n};1,3,1)\,.
\end{align}
Taking into account \thee decomposition \eqref{decomp}, \thee gauge
group is broken to $SU(3)^3$. Now, under $SU(3)^3$, \thee
surviving fields transform as
\begin{align}
  &SU(3)\times SU(3)\times SU(3)\,,\\
  &\left((3,\bar{3},1)+(\bar{3},1,3)+(1,3,\bar{3})\right)\,,
\end{align}
which correspond to \thee desired chiral
representations of \thee trinification group. Under $SU(3)_c\times SU(3)_L\times SU(3)_R$,
\thee quarks and leptons of \thee
first family transform as
\begin{align}
  q=\left(
      \begin{array}{ccc}
        d & u & h \\
        d & u & h \\
        d & u & h \\
      \end{array}
    \right)\sim(3,\bar{3},1)\,, q^c=\left(
                                             \begin{array}{ccc}
                                               d^c & d^c & d^c \\
                                               u^c & u^c & u^c \\
                                               h^c & h^c & h^c \\
                                             \end{array}
                                           \right)\sim(\bar{3},1,3)\,,
                                           \lambda=\left(
                                                     \begin{array}{ccc}
                                                       N & E^c & \text{v} \\
                                                       E & N^c & e \\
                                                       \text{v}^c & e^c & S \\
                                                     \end{array}
                                                   \right)\sim(1,3,\bar{3})\,,
\end{align}
respectively. Similarly, one obtains \thee matrices for \thee fermions of \thee
other two families.

It is worth noting that this theory can be upgraded to a two-loop
finite theory (for reviews see \cite{inspire2, inspire4,
heinemeyer18, MaZoup}) and give testable
predictions \cite{MaZoup}, too. Additionally, fuzzy orbifolds can be used
to break spontaneously \thee unification gauge group down to
MSSM and then to \thee $SU(3)_c\times U(1)_{em}$.

Summarizing this section let us emphasize \thee general picture of \thee
model that has been constructed. At very high-scale regime, we have an unbroken
renormalizable theory. After \thee spontaneous symmetry breaking, \thee resulting gauge theory is accompanied \by a finite tower of massive Kaluza-Klein modes.
Finally, \thee theory breaks down to an extension of MSSM in \thee low scale regime. Therefore, we conclude that fuzzy extra dimensions
can be used in constructing chiral, renormalizable and
phenomenologically viable field-theoretical models.

A natural extension of \thee above ideas \andd methods have been
reported in ref \cite{chatzi-stein-zoup} (see also
\cite{chatzi-stein-zoup2}), realized in \thee context of Matrix
Models (MM). At a fundamental level, \thee MMs introduced \by
Banks-Fischler-Shenker-Susskind (BFSS) \andd
Ishibashi-Kawai-Kitazawa-Tsuchiya (IKKT), are supposed to provide a
non-perturbative definition of M-theory \andd type IIB string theory
respectively \cite{ishibasi-kawai, banks}. On \thee other hand, MMs
are also useful laboratories for \thee study of structures which
could be relevant from a low-energy point of view. Indeed, they
generate a plethora of interesting solutions, corresponding to
strings, D-branes \andd their interactions \cite{ishibasi-kawai,
chepelev}, \as well \as to non-commutative/fuzzy spaces, such \as
fuzzy tori \andd spheres \cite{iso}. Such backgrounds naturally give
rise to non-abelian gauge theories. Therefore, it appears natural to
pose \thee question whether it is possible to construct
phenomenologically interesting particle physics models in this
framework \as well. In addition, an orbifold MM was proposed \by
Aoki-Iso-Suyama (AIS) in \cite{aoki} \as a particular projection of
\thee IKKT model, \andd it is directly related to \thee construction
described above in which fuzzy extra dimensions arise with
trinification gauge theory \cite{fuzzy}. By $\mathbb{Z}_3$ -
orbifolding, \thee original symmetry of \thee IKKT matrix model with
matrix size $3N\times 3N$ is generally reduced from $SO(9,1)\times
U(3N)$ to $SO(3,1) \times U(N)^3$. This model is chiral \andd has
$D=4$, $\mathcal{N}=1$ supersymmetry of Yang-Mills type \as well \as
an inhomogeneous supersymmetry specific to matrix models. The
$\mathbb{Z}_3$ - invariant fermion fields transform \as
bifundamental representations under \thee unbroken gauge symmetry
exactly \as in \thee constructions described above. In \thee future
we plan to extend further \thee studies initiated in refs
\cite{chatzi-stein-zoup, chatzi-stein-zoup2} in \thee context of
orbifolded IKKT models.\\

Our current interest is to continue in two directions. Given that \thee
two approaches discussed here led to \thee $\mathcal{N}=1$ trinification GUT $SU(3)^3$,
one plan is to examine \thee phenomenological consequences of these models.
The models are different in \thee details but certainly there exist a
certain common ground. Among others we plan to determine in both cases \thee
spectrum of \thee Dirac and Laplace operators in \thee extra dimensions and
use them to study \thee behaviour of \thee various couplings, including
\thee contributions of \thee massive Kaluza-Klein modes. These contributions
are infinite or finite in number, depending on whether \thee extra dimensions are
continuous or fuzzy, respectively. We should note that \thee spectrum of \thee Dirac
operator at least in \thee case of $SU(3)/U(1)\times U(1)$ is not known.

Another plan is to start with an abelian theory in ten dimensions and
with a simple reduction to obtain an $\mathcal{N}=(1,1)$ abelian theory in six
dimensions. Finally, reducing \thee latter theory over a fuzzy sphere,
possibly with Chern-Simons terms, to obtain a non-abelian gauge theory in
four dimensions provided with soft supersymmetry breaking terms. Recall that
\thee last feature was introduced \by hand in \thee realistic models
constructed in \thee fuzzy extra dimensions framework.
\\\\
\emph{Acknowledgement}
\\
\noindent This research is supported \by \thee Research Funding Program ARISTEIA, Higher Order Calculations and
Tools for High Energy Colliders, HOCTools and \thee ARISTEIA II,
Investigation of certain higher derivative term field theories and
gravity models (co-financed \by \thee European Union (European Social
Fund ESF) and Greek national funds through \thee Operational Program
Education and Lifelong Learning of \thee National Strategic Reference
Framework (NSRF)), \by \thee European Union's ITN programme
HIGGSTOOLS, as well as \by \thee Action MP1405 QSPACE from \thee European Cooperation in Science and Technology
(COST). One of \thee authors, G.Z., would like to thank \thee organizers and ITP-Heidelberg
for \thee warm and generous hospitality.

\end{document}